\documentclass[amsmath,amssymb,showpacs,preprint]{revtex4}
\usepackage{epsfig}
\usepackage{graphicx}

\begin{document}

\title{Exact solution of vacuum field equation in Finsler spacetime}

\author{Xin Li $^{1,2,3}$}
\email{lixin1981@cqu.edu.cn}
\author{Zhe Chang $^{2,3}$}
\email{changz@ihep.ac.cn}
\affiliation{$^1$Department of Physics, Chongqing University, Chongqing 401331, China\\
$^2$Institute of High Energy Physics, Chinese Academy of Sciences, 100049 Beijing, China\\
$^3$State Key Laboratory Theoretical Physics, Institute of Theoretical Physics, Chinese
Academy of Sciences, Beijing 100190, China}

\begin{abstract}
We suggest that the vacuum field equation in Finsler spacetime is equivalent to vanishing of Ricci scalar. Schwarzschild metric can be deduced from a solution of our field equation if the spacetime preserve spherical symmetry. Supposing spacetime to preserve the symmetry of ``Finslerian sphere",  we find a non-Riemannian exact solution of the Finslerian vacuum field equation. The solution is similar to the Schwarzschild metric. It reduces to Schwarzschild metric while the Finslerian parameter $\epsilon$ vanishes. It is proved that the Finslerian covariant derivative of the geometrical part of the gravitational field equation is conserved. The interior solution is also given. We get solutions of geodesic equation in such a Schwarzschild-like spacetime, and show that the geodesic equation returns to the counterpart in Newton's gravity at weak field approximation. The celestial observations give constraint on the Finslerian parameter $\epsilon<10^{-4}$. And the recent Michelson-Morley experiment requires $\epsilon<10^{-16}$. The counterpart of Birkhoff's theorem exist in Finslerian vacuum. It shows that the Finslerian gravitational field with the symmetry of ``Finslerian sphere" in vacuum must be static.
\end{abstract}

\maketitle
\section{Introduction}
In 1912, Einstein proposed his famous general relativity which gives the connection between Riemann geometry and gravitation. In general relativity, the effects of gravitation are ascribed to spacetime curvature instead of a force. In four dimensional spacetime, two solutions of the Einstein vacuum field equation are well known \cite{Wald}. These are the Schwarzschild solution which preserves spherical symmetry, and the Kerr solution which preserves axial symmetry. The Schwarzschild solution is of vital importance for general relativity. The physics of Schwarzschild solution is quite different from Newton's gravity. The success of general relativity attribute to the four classical tests \cite{Weinberg}. The predictions of the four classical tests directly come from the Schwarzschild solution. Most celestial bodies can be approximately treated as a sphere. Thus, the Schwarzschild solution is widely used in investigating the astronomical phenomena. However, the recent astronomical observations show that the gravitational field of galaxy clusters offset from its baryonic matters \cite{Shan}. It implies that the spherical symmetry may not preserve in the scale of galaxy clusters.

Finsler geometry \cite{Book by Bao} is a new geometry which involves Riemann geometry as its special case. S. S. Chern pointed out that Finsler geometry is just Riemann geometry without quadratic restriction, in his Notices of AMS. The symmetry of spacetime is described by the isometric group. The generators of isometric group are directly connected with the Killing vectors. It is well known that the isometric group is a Lie group in Riemannian manifold. This fact also holds in Finslerian manifold \cite{Deng}. Generally, Finsler spacetime admits less Killing vectors than Riemann spacetime does \cite{Finsler PF}. The numbers of independent Killing vectors of an $n$ dimensional non-Riemannian Finsler spacetime should no more than $\frac{n(n-1)}{2}+1$ \cite{Wang}. The causal structure of Finsler spacetime is determined by the vanish of Finslerian length \cite{Pfeifer1}. And the speed of light is modified. It has been shown that the translation symmetry is preserved in flat Finsler spacetime \cite{Finsler PF}. Thus, the energy and momentum are well defined in Finsler spacetime. In flat Finsler spacetime, inertial motion preserves the Finslerian length and admits a modified dispersion relation.

Lorentz invariance (LI) is one of the foundations of the Standard
model of particle physics. Of course, it is very interesting to test
the fate of the LI both on experiments and theories. The theoretical approach of investigating the LI violation is studying the possible spacetime symmetry, and erecting some counterparts of special relativity. Recently, there are a few counterparts of special relativity. The first one is doubly special relativity (DSR)\cite{Amelino1,Amelino2,Amelino3}. In DSR, the Planck-scale effects have been taken into account by introducing an invariant Planckin parameter. Together with the speed of light, DSR has two invariant parameters. The second one is very special relativity (VSR) \cite{Coleman}. Coleman and Glashow have set up a perturbative framework for investigating possible departures of local quantum field theory from LI. The symmetry group of VSR is some certain subgroups of Poincare group, which contains the spacetime translations and proper subgroups of Lorentz transformations.
Gibbons {\it et al}. \cite{Gibbons} have pointed out that Glashow's VSR is Finsler Geometry. Generally, the flat Finsler spacetime breaks the Lorentz symmetry. Thus, it is a possible mechanism of Lorentz violation \cite{Kostelecky}.

The standard cosmological model, i.e., the $\Lambda$CDM model \cite{Sahni,Padmanabhan} has been well established. It is consistent with several precise astronomical observations that involve Wilkinson Microwave Anisotropy Probe (WMAP) \cite{Komatsu}, Planck satellite \cite{Planck1}, Supernovae Cosmology Project \cite{Suzuki}. One of the most important and basic assumptions of the $\Lambda$CDM model states that the universe is homogeneous and isotropic on large scales. However, such a principle faces challenges \cite{Perivolaropoulos}. The Union2 SnIa data hint that the universe has a preferred direction $(l,b)=(309^\circ,18^\circ)$ in galactic coordinate system \cite{Antoniou}. Toward this direction, the universe has the maximum expansion velocity. Astronomical observations \cite{Watkins} found that the dipole moment of the peculiar velocity field on the direction $(l,b)=(287^\circ\pm9^\circ,8^\circ\pm6^\circ)$ in the scale of $50h^{-1}\rm Mpc$ has a magnitude $407\pm81 \rm km\cdot s^{-1}$. This peculiar velocity is much larger than the value $110 \rm km\cdot s^{-1}$ given by WMAP5 \cite{WMAP5}. The recent released data of Planck Collaboration show deviations from isotropy with a level of significance ($\sim3\sigma$) \cite{Planck2}. Planck Collaboration confirms asymmetry of the power spectrums between two preferred opposite hemispheres. These facts hint that the universe may have a preferred direction.

Many models have been proposed to resolve the asymmetric anomaly of the astronomical observations. An incomplete and succinct list includes: an imperfect fluid dark energy \cite{Koivisto1}, local void scenario \cite{Alexander,Garcia}, noncommutative spacetime effect \cite{Akofor}, anisotropic curvature in cosmology \cite{Koivisto2}, and Finsler gravity scenario \cite{Chang}. Instead of Minkowski spacetime, Finsler spacetime involves a preferred direction. It is a natural candidate for describing the cosmological anisotropy.

Both the $\Lambda$CDM model and particle standard model require no variation of fundamental physical constants in principle, such as fine structure constant $\alpha=e^2/\hbar c$. Recently, the observations on quasar absorption spectra show that the fine structure constant varies at cosmological scale \cite{Webb}. Furthermore, in high redshift region ($z>1.6$), they have shown that the variation of $\alpha$ is well represented by an angular dipole model pointing in the direction $(l,b)=(330^\circ,-15^\circ)$ with level of significance ($\sim4.2\sigma$). Mariano and Perivolaropoulos \cite{Mariano} have shown that the dipole of $\alpha$ is anomalously aligned with corresponding dark energy dipole obtained through the Union2 sample. One direct reason of the variation of $\alpha$ is the variation of the speed of light. Finsler geometry, as a natural tool for investigating both the cosmological preferred direction and Lorentz violations, may be also referred as a natural framework to describe the dipole structure of fine structure constant. Models \cite{Cahng:2013zwa,Chang:2013vla,Chang:2013lxa} based on Finsler spacetime have been developed to study the cosmological preferred direction.

The counterpart of special relativity has been established in flat Finsler spacetime. However, up to now, Finslerian gravity is still to be completed. There are types of gravitational field equation in Finsler spacetime \cite{Li Berwald,Miron,Rutz1,Vacaru}. Models have been built to study the gravitational theory that constructed on Finsler spacetime. Stavrinos {\it et al}. \cite{Stavrinos,Chang:2013vla} used the method of osculating Riemannian space to study the cosmological anisotropy in Finsler spacetime. Vacaru {\it et al}. studied high dimensional gravity in Finsler spacetime \cite{Vacaru ref}. Pfeifer {\it et al}. \cite{Pfeifer} have constructed gravitational dynamics for Finsler spacetimes in terms of an action integral on the unit tangent bundle. However, these equations (or models) are not equivalent to each other. Here are the reasons. It is well known that there is only a torsion free connection-the Christoffel connection in Riemann geometry. However, there are types of connection in Finsler geometry. Therefore, the covariant derivatives that depend on the connection are different. The Finslerian length element $F$ is constructed on a tangent bundle \cite{Book by Bao}. Thus, the gravitational field equation should be constructed on the tangent bundle in principle. However, the corresponded energy-momentum tensor, which should be constructed on the tangent bundle, is rather obscure in physics.

The analogy between geodesic deviation equations in Finsler spacetime and Riemann spacetime gives the vacuum field equation in Finsler gravity \cite{Finsler Bullet}. It is the vanishing of Ricci scalar. The vanishing of the Ricci scalar implies that the geodesic rays are parallel to each other. The geometric invariant property of Ricci scalar implies that the vacuum field equation is insensitive to the connection, which is an essential physical requirement. In this paper, we present an exact solution of the vacuum field equation in Finsler spacetime. The interior solution is also shown.

This paper is organized as follows. Sec.II is divided into two subsections. In subsection A, we briefly introduce some basic geometric objects in Finsler geometry. And we show the symmetry of flat Finsler spacetime. The spherical symmetry can be represented by metric of Riemannian sphere. In subsection B, we introduce ``Finslerian sphere" that is a counterpart of Riemannian sphere. Sec.III is divided into four subsections. In subsection A, we introduce vacuum field equation in Finsler spacetime by analogizing geodesic deviation equation in Finsler spacetime and one in Riemann spacetime. In subsection B, we present an exact solution of the vacuum field equation in Finsler spacetime. In subsection C, we investigate the Newtonian limit of our solution. In subsection D, we show boundary conditions of the vacuum field equation. It will distinguish Schwarzschild solution from our solution. In Sec.IV, we propose a gravitational field equation with source. We prove that the Finslerian covariant derivative of the geometrical part of the gravitational field equation is conserved. Then, an interior solution of gravitational field equation is shown. In Sec.V, we investigate experimental constraint on the Schwarzschild-like spacetime that is given in Sec. III. Particle motion is shown in subsection A. We get three solutions of geodesic equations. In solar system, experimental constraints on our Finslerian parameter is given in subsection B. In Sec.VI, we show counterpart of the Birkhoff's theorem. It states that Finslerian gravitational field with the symmetry of ``Finslerian sphere" in vacuum must be static. Conclusions and remarks are given in Sec. VII.

\section{Symmetry of Finsler spacetime}
\subsection{Killing vectors}
Instead of defining an inner product structure over the tangent bundle in Riemann geometry, Finsler geometry is based on
the so called Finsler structure $F$ with the property
$F(x,\lambda y)=\lambda F(x,y)$ for all $\lambda>0$, where $x\in M$ represents position
and $y\equiv\frac{dx}{d\tau}$ represents velocity. The Finslerian metric is given as\cite{Book
by Bao}
\begin{equation}
g_{\mu\nu}\equiv\frac{\partial}{\partial
y^\mu}\frac{\partial}{\partial y^\nu}\left(\frac{1}{2}F^2\right).
\end{equation}
In physics, the Finsler structure $F$ is not positive-definite at every point of Finsler manifold. We focus on investigating Finsler spacetime with Lorentz signature. A positive, zero and negative $F$ correspond to time-like, null and space-like curves, respectively. For massless particles, the stipulation is $F=0$.
Two types of Finsler space should be noticed. One is the Riemann space. A Finslerian metric is said to be Riemannian, if $F^2$ is quadratic in $y$. Another is Randers spacetime \cite{Randers}. It is given as
\begin{equation}\label{Randers form}
F(x,y)\equiv \alpha(x,y)+\beta(x,y),
\end{equation} where
\begin{eqnarray}
\alpha (x,y)&\equiv&\sqrt{\tilde{a}_{\mu\nu}(x)y^\mu y^\nu},\\
\beta(x,y)&\equiv& \tilde{b}_\mu(x)y^\mu,
\end{eqnarray} and $\tilde{a}_{ij}$ is Riemannian metric. Throughout this paper, the indices are lowered and raised by $g_{\mu\nu}$ and its inverse matrix $g^{\mu\nu}$. And the objects that decorate with a tilde are lowered and raised by $\tilde{a}_{\mu\nu}$ and its inverse matrix $\tilde{a}^{\mu\nu}$.

To investigate the Killing vectors, we should construct the isometric
transformations of Finsler structure. It is convenient to
discuss the isometric transformations under an infinitesimal
coordinate transformation for $x$
\begin{equation}
\label{coordinate tran}
\bar{x}^\mu=x^\mu+\epsilon \tilde{V}^\mu,
\end{equation}
together with a corresponding transformation for $y$
\begin{equation}
\label{coordinate tran1}
\bar{y}^\mu=y^\mu+\epsilon\frac{\partial \tilde{V}^\mu}{\partial x^\nu}y^\nu,
\end{equation}
where $|\epsilon|\ll1$.
Under the coordinate transformation (\ref{coordinate tran}) and (\ref{coordinate tran1}), to first order in $|\epsilon|$, we obtain the expansion of the Finsler structure,
\begin{equation}
\label{coordinate tran F}
\bar{F}(\bar{x},\bar{y})=\bar{F}(x,y)+\epsilon \tilde{V}^\mu\frac{\partial F}{\partial x^\mu}+\epsilon y^\nu\frac{\partial \tilde{V}^\mu}{\partial x^\nu}\frac{\partial F}{\partial y^\mu},
\end{equation}
where $\bar{F}(\bar{x},\bar{y})$ should equal to $F(x,y)$.
Under the transformation (\ref{coordinate tran}) and (\ref{coordinate tran1}), a Finsler structure is called isometry if and only if
\begin{equation}
F(x,y)=\bar{F}(x,y).
\end{equation}
Deducing from the (\ref{coordinate tran F}), we obtain the Killing equation $K_V(F)$ in Finsler space
\begin{equation}
\label{killing F}
K_V(F)\equiv \tilde{V}^\mu\frac{\partial F}{\partial x^\mu}+y^\nu\frac{\partial \tilde{V}^\mu}{\partial x^\nu}\frac{\partial F}{\partial y^\mu}=0.
\end{equation}

Plugging the length element of Randers spacetime (\ref{Randers form}) into the Killing equation (\ref{killing F1}), and noticing that the rational and irrational parts of Killing equation are independent, we obtain that
\begin{eqnarray}
\label{killing F1}
\tilde{V}_{\mu|\nu}+\tilde{V}_{\nu|\mu}&=&0,\\
\label{killing F2}
\tilde{V}^\mu\frac{\partial \tilde{b}_\nu}{\partial x^\mu}+\tilde{b}_\mu\frac{\partial \tilde{V}^\mu}{\partial x^\nu}&=&0,
\end{eqnarray}
where $``|"$ denotes the covariant derivative with respect to the Riemannian metric $\alpha$. These equations (\ref{killing F1},\ref{killing F2}) are equivalent to the statements
that $L_V \alpha = 0$ and $L_V \beta = 0$, respectively, where $\alpha$ is the obvious Riemannian metric and $\beta$ is the one form. Here $L_V$ is the Lie derivative
along V. It is obvious that Killing equation (\ref{killing F1}) is the same as the Riemannian one. But, the other Killing equation (\ref{killing F2}) obtained constrains the first one (\ref{killing F1}). Thus, the number of independent killing vectors in Randers-Finsler spacetime (\ref{Randers form}) is less than the one in Riemannian spacetime $\alpha$ \cite{Finsler PF}.

The geodesic equation for Finsler manifold is given as
\begin{equation}
\label{geodesic}
\frac{d^2x^\mu}{d\tau^2}+2G^\mu=0,
\end{equation}
where
\begin{equation}
\label{geodesic spray}
G^\mu=\frac{1}{4}g^{\mu\nu}\left(\frac{\partial^2 F^2}{\partial x^\lambda \partial y^\nu}y^\lambda-\frac{\partial F^2}{\partial x^\nu}\right)
\end{equation} is called geodesic spray coefficients. It can be proved from the geodesic equation (\ref{geodesic}) that the Finslerian structure $F\left(x,\frac{dx}{d\tau}\right)$ is constant along the geodesic.

A Finsler metric is said to be locally Minkowskian if at every point, there is a local coordinate system, such that $F = F (y)$
is independent of the position $x$ \cite{Book by Bao}. It can be proved that all types of curvature tensors vanish in locally Minkowskian spacetime. Thus, the locally Minkowskian spacetime is flat Finsler spacetime. The momentum in flat Finsler spacetime are defined as $p^\mu\equiv m\frac{dx^\mu}{d\tau}$. Since the Finsler structure $F$ does not depend on $x$, it is clear from the geodesic equation that $\frac{dp^\mu}{d\tau}=0$. Therefore, momentum $p^\mu$ are conserved in flat Finsler spacetime. The dispersion relation of flat Finsler spacetime is given as
\begin{equation}\label{MDR}
\eta_{\mu\nu}(p)p^\mu p^\nu=m^2.
\end{equation}
One should notice that the $\eta_{\mu\nu}(p)$ is not a constant in flat Finsler spacetime. For example, in flat Randers spacetime, its dispersion relation is given as
\begin{equation}
\sqrt{\tilde{\eta}_{\mu\nu}p^\mu p^\nu}+\tilde{b}_\rho p^\rho=m,
\end{equation}
where $\tilde{\eta}_{\mu\nu}$ is Minkowski metric. The flat Finsler spacetime not only give modified dispersion relation that plays the role of testing Lorentz invariance, but also is taken as boundary condition to solve the gravitational field equation in Finsler spacetime.

\subsection{Two dimensional Finsler space with constant flag curvature}
In general relativity, the Schwarzschild metric preserves spherical symmetry. In Riemann geometry, at a fixed radial coordinate $r$, if the metric is of the form
\begin{equation}\label{Riemann sphere}
F_{\rm RS}=\sqrt{(y^\theta)^2+\sin\theta (y^\varphi)^2},
\end{equation}
then, we call it possesses spherical symmetry. It is clear that the Riemann space with metric (\ref{Riemann sphere}) has constant curvature, and it equals 1. By using Killing equation (\ref{killing F}), one can find that the metric (\ref{Riemann sphere}) has three independent Killing vectors. In this paper, we want to find a Schwarzschild-like solution in Finsler spacetime. In Finsler geometry, the counterpart of spherical symmetry is ``Finslerian sphere". In order to represent feature of most celestial bodies, the ``Finslerian sphere" should look like a sphere. In mathematics, it should topologically equivalent to a sphere. Also, the ``Finslerian sphere" should preserve as much symmetry as it can. The theorem proved in reference \cite{Wang} tells that the two dimensional Finsler space has only one independent Killing vector. So does the ``Finslerian sphere". In Finsler geometry, the generalization of Riemannian sectional curvature is flag curvature. The constant flag curvature is equivalent to the constant Ricci scalar. Thus, the counterpart of Riemannian sphere, the ``Finslerian sphere" should have constant flag curvature. In the followings, we present a specific example of ``Finslerian sphere".

Bao {\it et al}. \cite{Bao} have given a completely classification of Randers-Finsler space \cite{Randers} with constant flag curvature. A two dimensional Randers-Finsler space with constant positive flag curvature $\lambda=1$ is given as
\begin{equation}\label{Randers 1}
F_{\rm FS}=\frac{\sqrt{(1-\epsilon^2\sin^2\theta)y^\theta y^\theta+\sin^2\theta y^\varphi y^\varphi}}{1-\epsilon^2\sin^2\theta}-\frac{\epsilon\sin^2\theta y^\varphi}{1-\epsilon^2\sin^2\theta},
\end{equation}
where $0\leq\epsilon<1$. It is obvious that the metric (\ref{Randers 1}) returns to Riemannian sphere while $\epsilon=0$. And the metric (\ref{Randers 1}) is non-reversible for $\varphi\rightarrow-\varphi$. The Randers-Finsler space (\ref{Randers 1}) has two close geometrically distinct closed geodesics \cite{Rademacher} if $\epsilon$ is irrational. The two geodesics locate at $\theta=\frac{\pi}{2}$ with length $L_{\pm}=2\pi(1\pm\epsilon)^{-1}$. This fact can be proved by plugging the metric (\ref{Randers 1}) into the formula of geodesic equation (\ref{geodesic}). Then, one can find that $\theta=\frac{\pi}{2}$ and $\varphi=u\tau+v$ ($u,v$ are integral constants) are the solutions of geodesic equation. The Randers-Finsler space (\ref{Randers 1}) is homotopy equivalent to the two dimensional sphere \cite{Rademacher}.


In terms of Busemann-Hausdorff volumm form, the volume of a close Randers-Finsler surface $F=\sqrt{a_{ij}(x)y^iy^j}+b_i(x)y^i$ is given as \cite{Book Shen}
\begin{equation}\label{Busemann volume}
Vol_F=\int (1-(a^{ij}b_i b_j))^{\frac{3}{2}}\sqrt{\rm det(a_{ij})}dx^1\wedge dx^2.
\end{equation}
Plugging the Randers metric (\ref{Randers 1}) into the formula (\ref{Busemann volume}), we obtain that its volume is $4\pi$, which is the same with unit Riemannian sphere.

\section{The exact solution of vacuum field equation}
\subsection{Vacuum field equation}
In this paper, we introduce the vacuum field equation in the way first discussed by Pirani \cite{Pirani, Rutz}. In Newton's theory of gravity, the equation of motion of a test particle is given as
\begin{equation}
\label{dynamic Newton}
\frac{d^2x^i}{d\tau^2}=-\eta^{ij}\frac{\partial \phi}{\partial x^i},
\end{equation}
where $\phi=\phi(x)$ is the gravitational potential and $\eta^{ij}=$ \textmd{diag}(+1,+1,+1) is the Euclidean metric. For an infinitesimal transformation $x^i\rightarrow x^i+\epsilon\xi^i$($|\epsilon|\ll1$), the equation (\ref{dynamic Newton}) becomes, to first order of $\epsilon$,
\begin{equation}
\label{dynamic Newton1}
\frac{d^2x^i}{d\tau^2}+\epsilon\frac{d^2\xi^i}{d\tau^2}=-\eta^{ij}\frac{\partial \phi}{\partial x^i}-\epsilon\eta^{ij}\xi^k\frac{\partial^2\phi}{\partial x^j\partial x^k}.
\end{equation}
Combining equations (\ref{dynamic Newton}) and (\ref{dynamic Newton1}), we obtain
\begin{equation}
\frac{d^2\xi^i}{d\tau^2}=\eta^{ij}\xi^k\frac{\partial^2\phi}{\partial x^j\partial x^k}\equiv\xi^kH^i_{~k}.
\end{equation}
For the vacuum field equation, one has $H^i_{~i}=\bigtriangledown^2\phi=0$.

In general relativity, the geodesic deviation gives a similar equation
\begin{equation}
\frac{D^2\xi^\mu}{D\tau^2}=\xi^\nu \tilde{R}^\mu_{~\nu},
\end{equation}
where $\tilde{R}^\mu_{~\nu}=\tilde{R}^{~\mu}_{\lambda~\nu\rho}\frac{dx^\lambda}{d\tau}\frac{dx^\rho}{d\tau}$. Here, $\tilde{R}^{~\mu}_{\lambda~\nu\rho}$ is the Riemannian curvature tensor. $D$ denotes the covariant derivative along the curve $x^\mu(t)$. The vacuum field equation in general relativity gives $\tilde{R}^{~\lambda}_{\mu~\lambda\nu}=0$ \cite{Weinberg}. This implies that the tensor $\tilde{R}^\mu_{~\nu}$ is also traceless, $\tilde{R}\equiv\tilde{R}^\mu_{~\mu}=0$.

In Finsler spacetime, the geodesic deviation yields \cite{Book by Bao}
\begin{equation}
\frac{D^2\xi^\mu}{D\tau^2}=\xi^\nu R^\mu_{~\nu},
\end{equation}
where $R^\mu_{~\nu}=R^{~\mu}_{\lambda~\nu\rho}\frac{dx^\lambda}{d\tau}\frac{dx^\rho}{d\tau}$. Here, $R^{~\mu}_{\lambda~\nu\rho}$ is Finsler curvature tensor \cite{Book by Bao}, $D$ denotes covariant derivative $\frac{D\xi^\mu}{D\tau}=\frac{d\xi^\mu}{d\tau}+\xi^\nu\frac{dx^\lambda}{d\tau}\Gamma^\mu_{~\nu\lambda}(x,\frac{dx}{d\tau})$. Since the vacuum field equations of Newton's gravity and general relativity are of similar forms, we may assume that vacuum field equation in Finsler spacetime faces similar requirements as in the case of Netwon's gravity and general relativity. It implies that the tensor $R^\mu_{~\nu}$ in Finsler geodesic deviation equation should be traceless, $R^\mu_{~\mu}=0$. Since the Riemann curvature tensor $R^{~\mu}_{\lambda~\nu\rho}$ does not depend on $\frac{dx}{d\tau}$, the vanish of $R^\mu_{~\mu}$ equals to the vanish of $R^{~\mu}_{\lambda~\mu\rho}$ which is just the vacuum field equation in general relativity.

We have proved that the analogy of the geodesic deviation equation is valid at least in a Finsler spacetime of Berwald type \cite{Finsler DM}. We assume that this analogy still holds its validity in a general Finsler spacetime.
In Finsler geometry, there is a geometrical invariant --- the Ricci scalar $Ric$ . It is of the form \cite{Book by Bao}
\begin{equation}\label{Ricci scalar}
Ric\equiv R^\mu_{~\mu}=\frac{1}{F^2}\left(2\frac{\partial G^\mu}{\partial x^\mu}-y^\lambda\frac{\partial^2 G^\mu}{\partial x^\lambda\partial y^\mu}+2G^\lambda\frac{\partial^2 G^\mu}{\partial y^\lambda\partial y^\mu}-\frac{\partial G^\mu}{\partial y^\lambda}\frac{\partial G^\lambda}{\partial y^\mu}\right).
\end{equation}
The Ricci scalar depends only on the Finsler structure $F$ and is insensitive to the connection.
For a tangent plane $\Pi\subset T_xM$ and a non-zero vector $y\in T_xM$, the flag curvature is defined as
\begin{equation}
\label{flag curvature}
K(\Pi,y)\equiv\frac{g_{\lambda\mu}R^\mu_{~\nu}u^\nu u^\lambda}{F^2g_{\rho\theta}u^\rho u^\theta-(g_{\sigma\kappa}y^\sigma u^\kappa)^2},
\end{equation}
where $u\in\Pi$. The flag curvature is a geometrical invariant and a generalization of the sectional curvature in Riemannian geometry. The Ricci scalar $Ric$ is the trace of $R^\mu_{~\nu}$, which is the predecessor of flag curvature. Thus the value of Ricci scalar $Ric$ is invariant under the coordinate transformation.

Furthermore, the significance of the Ricci scalar $Ric$ is very clear. It plays an important role in the geodesic deviation equation \cite{Finsler GW,Finsler MOND,Book by Bao}. The vanishing of the Ricci scalar $Ric$ implies that the geodesic rays are parallel to each other. It means that it is vacuum outside the gravitational source.

Therefore, it is reasonable to believe that the gravitational vacuum field equation in Finsler geometry has its essence in $Ric=0$. Pfeifer and Wohlfarth \cite{Pfeifer} have constructed gravitational dynamics for Finsler spacetime in terms of an action integral on the unit tangent bundle. Their results show that the gravitational field equation in Finsler spacetime is given as
\begin{equation}\label{Gravitation eq Pfeifer}
S-6Ric+2g^{\mu\nu}\big(\nabla_\mu S_\nu+S_\mu S_\nu+\partial_{y^\mu}\nabla S_\nu\big)=-4\pi G T.
\end{equation}
The $S_\mu$-terms can be written as $S_\mu=y^{\nu}P_{\nu~\lambda \mu}^{~\lambda}/F$, where $P_{\nu~\lambda \mu}^{~\lambda}$ are the coefficients of the cross basis $dx \wedge \frac{\delta y}{F}$ \cite{Book by Bao}. Accordingly, the energy-momentum tensor can also be divided into two parts in terms of the basis of $dx\wedge dx$ and $dx \wedge \frac{\delta y}{F}$, respectively. Thus, the $S_\mu$-terms contribute to the energy-momentum tensor that belong to the basis $dx \wedge \frac{\delta y}{F}$. The vacuum field equation constructed by Pfeifer and Wohlfarth implies that each coefficients of different basis should vanish respectively. Thus, the stipulation $Ric=0$ here is compatible with Pfeifer and Wohlfarth's results of gravitational field equation.

\subsection{Vacuum solution}
Here, we propose an ansatz that the Finsler structure is of the form
\begin{equation}\label{Schwarz like}
F^2=B(r)y^ty^t-A(r)y^r y^r-r^2\bar{F}^2(\theta,\varphi,y^\theta,y^\varphi).
\end{equation}
Then, the Finsler metric can be derived as
\begin{eqnarray}
g_{\mu\nu}&=&{\rm diag}(B,-A,-r^2\bar{g}_{ij}),\\
g^{\mu\nu}&=&{\rm diag}(B^{-1},-A^{-1},-r^{-2}\bar{g}^{ij}),
\end{eqnarray}
where $\bar{g}_{ij}$ and its reverse are the metric that derived from $\bar{F}$ and the index $i,j$ run over angular coordinate $\theta,\varphi$.

Plugging the Finsler structure (\ref{Schwarz like}) into the formula (\ref{geodesic spray}), we obtain that
\begin{eqnarray}\label{geodesic spray t}
G^t&=&\frac{B'}{2B}y^t y^r,\\
\label{geodesic spray r}
G^r&=&\frac{A'}{4A}y^r y^r+\frac{B'}{4A}y^t y^t-\frac{r}{2A}\bar{F}^2,\\
\label{geodesic spray theta}
G^\theta&=&\frac{1}{r}y^\theta y^r+\bar{G}^\theta,\\
\label{geodesic spray phi}
G^\varphi&=&\frac{1}{r}y^\varphi y^r+\bar{G}^\varphi,
\end{eqnarray}
where the prime denotes the derivative with respect to $r$, and the $\bar{G}$ is the geodesic spray coefficients derived by $\bar{F}$.
Plugging the geodesic coefficients (\ref{geodesic spray t},\ref{geodesic spray r},\ref{geodesic spray theta},\ref{geodesic spray phi}) into the formula of Ricci scalar (\ref{Ricci scalar}), we obtain that
\begin{eqnarray}
F^2Ric=&&\left[\frac{B''}{2A}-\frac{B'}{4A}\left(\frac{A'}{A}+\frac{B'}{B}\right)+\frac{B'}{r A}\right]y^ty^t\nonumber\\
&+&\left[-\frac{B''}{2B}+\frac{B'}{4B}\left(\frac{A'}{A}+\frac{B'}{B}\right)+\frac{A'}{r A}\right]y^r y^r\nonumber\\
\label{Ricci scalar1}
&+&\left[\bar{R}ic-\frac{1}{A}+\frac{r}{2A}\left(\frac{A'}{A}-\frac{B'}{B}\right)\right]\bar{F}^2~,
\end{eqnarray}
where $\bar{R}ic$ denotes the Ricci scalar of Finsler structure $\bar{F}$.
Since $\bar{F}$ is independent of $y^t$ and $y^r$, the vanish of Ricci scalar implies that the term in each square bracket of equation (\ref{Ricci scalar1}) should vanish respectively. These equations are given as
\begin{eqnarray}\label{eq1}
0&=&\frac{B''}{2A}-\frac{B'}{4A}\left(\frac{A'}{A}+\frac{B'}{B}\right)+\frac{B'}{r A},\\
\label{eq2}
0&=&-\frac{B''}{2B}+\frac{B'}{4B}\left(\frac{A'}{A}+\frac{B'}{B}\right)+\frac{A'}{r A},\\
\label{eq3}
0&=&\bar{R}ic-\frac{1}{A}+\frac{r}{2A}\left(\frac{A'}{A}-\frac{B'}{B}\right)~.
\end{eqnarray}
Noticing that $\bar{R}ic$ is independent of $r$, thus the equation (\ref{eq3}) is satisfied if and only if $\bar{R}ic$ equals to constant. It means that the two dimensional Finsler space $\bar{F}$ is a constant flag curvature space. The flag curvature is a generalization of sectional curvature in Riemann geometry. Here, we label the constant flag curvature to be $\lambda$. Therefore, $\bar{R}ic=\lambda$. The equations (\ref{eq1},\ref{eq2},\ref{eq3}) are similar to the Schwarzschild vacuum field equation in general relativity. The solutions of equations (\ref{eq1},\ref{eq2},\ref{eq3}) are given as
\begin{eqnarray}
\label{solution B1}
B&=&a\lambda+\frac{b}{r},\\
\label{solution A1}
A&=&\left(\lambda+\frac{b}{ra}\right)^{-1},
\end{eqnarray}
where $a$ and $b$ are integral constants.

\subsection{The Newtonian limit}
In the above subsection, we have obtained the vacuum field solution in Finsler spacetime. The integral constants of solutions (\ref{solution B1},\ref{solution A1}) should be determined by specific boundary conditions which is given by physical requirement. Here, we require that the solutions should return to Newton's gravity in weak field approximation \cite{Weinberg}.  In order to compare with Newton's gravity, we only consider the radial motion of particles. Plugging the solutions (\ref{solution B1},\ref{solution A1}) into the geodesic coefficients (\ref{geodesic spray t},\ref{geodesic spray r}), and noticing that the velocity of particle $\frac{dr}{dt}$ is small, we obtain the geodesic equations
\begin{eqnarray}\label{Newton1}
\frac{d^2t}{d\tau^2}=0,\\
\label{Newton2}
\frac{d^2r}{d\tau^2}-\frac{b\lambda}{2r^2}\left(\frac{dt}{d\tau}\right)^2=0.
\end{eqnarray}
Combining the geodesic equations (\ref{Newton1},\ref{Newton2}), we obtain that
\begin{equation}\label{Newton3}
\frac{d^2r}{dt^2}=\frac{b\lambda}{2r^2}.
\end{equation}
Comparing the equation (\ref{Newton3}) with Newton's gravity, we conclude that
\begin{equation}
\label{Newtonian limit}
b\lambda=-2GM,
\end{equation}
where $M$ denotes the total mass of gravitational source.

\subsection{Boundary conditions}
The solution of vacuum field equation $Ric=0$ gives specific form of function $B(r)$ and $A(r)$, and requires the two dimensional subspace $\bar{F}$ to be a constant curvature space. Two integral constants $a$ and $b$ and the specific form of subspace $\bar{F}$ need to be determined by boundary conditions. These boundary conditions are given by physical requirements. In subsection C, we get $b=-2GM/\lambda$ by the Newtonian limit. In the following section, we will show that the interior solution is consistent with the exterior solution at the boundary of the gravitational source if $a=1/\lambda$. The value of constant curvature $\lambda$ can be set to 1 by redefinition of the curve parameter $\tau$. Now, one boundary condition is left to determine the specific form of subspace $\bar{F}$.

In general relativity, the Schwarzschild metric returns to the Minkowski metric if $r\rightarrow\infty$. It means that the spacetime is Minkowski in the absent of gravity. The Finsler spacetime (\ref{Schwarz like}) in the absent of gravity is another physical boundary condition. If $r\rightarrow\infty$ or $M=0$, the Finsler spacetime (\ref{Schwarz like}) reduces to Minkowski spacetime, then our solution of vacuum field equation is no other than the Schwarzschild solution. This fact implies that Finsler geometry is a generation of Riemann geometry and Finslerian gravity involves the physical contents of general relativity. If $r\rightarrow\infty$ or $M=0$, the Finsler spacetime (\ref{Schwarz like}) violates Lorentz symmetry, then we get a Finslerian solution of vacuum field equation. For example, $\bar{F}$ is of the form $F_{\rm FS}$ (\ref{Randers 1}).

\section{Interior solution}
It is well known that the Schwarzschild spacetime has interior solution, which can deduce the famous Oppenheimer-Volkoff equation. The interior behavior of Finsler spacetime (\ref{Schwarz like}) is worth investigating. However, as we mentioned in the introduction, there are obstructions in constructing gravitational field equation in Finsler spacetime. In this section, we will show that there is a self-consistent gravitational field equation in Finsler spacetime (\ref{Schwarz like}).

The notion of Ricci tensor in Finsler geometry was first introduced by Akbar-Zadeh\cite{Akbar}
\begin{equation}\label{Ricci tensor}
Ric_{\mu\nu}=\frac{\partial^2\left(\frac{1}{2}F^2 Ric\right)}{\partial y^\mu\partial y^\nu}.
\end{equation}
And the scalar curvature in Finsler geometry is given as $S=g^{\mu\nu}Ric_{\mu\nu}$. Here, we define the modified Einstein tensor in Finsler spacetime
\begin{equation}\label{Einstein tensor}
G_{\mu\nu}\equiv Ric_{\mu\nu}-\frac{1}{2}g_{\mu\nu}S.
\end{equation}
Plugging the equation of Ricci scalar (\ref{Ricci scalar1}) into the formula (\ref{Einstein tensor}), and noticing that $\bar{F}$ is two dimensional Finsler spacetime with constant flag curvature $\lambda$, we obtain
\begin{eqnarray}\label{G t}
G^t_t&=&\frac{A'}{rA^2}-\frac{1}{r^2A}+\frac{\lambda}{r^2},\\
\label{G r}
G^r_r&=&-\frac{B'}{rAB}-\frac{1}{r^2A}+\frac{\lambda}{r^2},\\
\label{G theta}
G^\theta_\theta=G^\varphi_\varphi&=&-\frac{B''}{2AB}-\frac{B'}{2rAB}+\frac{A'}{2rA^2}+\frac{B'}{4AB}\left(\frac{A'}{A}+\frac{B'}{B}\right).
\end{eqnarray}

Next, we investigate the covariant conserve properties of the tensor $G^\mu_\nu$. The covariant derivative of $G^\mu_\nu$ in Finsler spacetime is given as \cite{Book by Bao}
\begin{equation}\label{covariant der}
G^\mu_{\nu~|\mu}=\frac{\delta}{\delta x^\mu}G^\mu_\nu+\Gamma^\mu_{\mu\rho}G^\rho_\nu-\Gamma^\rho_{\mu\nu}G^\mu_\rho,
\end{equation}
where
\begin{equation}\label{delta der}
\frac{\delta}{\delta x^\mu}=\frac{\partial}{\partial x^\mu}-\frac{\partial G^\rho}{\partial y^\mu}\frac{\partial}{\partial y^\rho},
\end{equation}
and $\Gamma^\mu_{\mu\rho}$ is the Chern connection. Here, we have used `$|$' to denote the covariant derivative. The form of covariant derivative (\ref{covariant der}) and `$\delta$'-derivative (\ref{delta der}) are well defined such that they transform as tensor under a coordinate change in Finsler spacetime \cite{Book by Bao}. The Chern connection can be expressed in terms of geodesic spray coefficients $G^\mu$ and Cartan connection $A_{\lambda\mu\nu}\equiv\frac{F}{4}\frac{\partial}{\partial y^\lambda}\frac{\partial}{\partial y^\mu}\frac{\partial}{\partial y^\nu}(F^2)$
\begin{equation}\label{chern connection}
\Gamma^\rho_{\mu\nu}=\frac{\partial^2 G^\rho}{\partial y^\mu \partial y^\nu}-A^\rho_{\mu\nu|\kappa}\frac{y^\kappa}{F}.
\end{equation}
Noticing that the modified Einstein tensor $G^\mu_\nu$ only depend on $r$ and do not have $y$-dependence, and Cartan tensor $A^\rho_{\mu\nu}=A^i_{jk}$ (index $i,j,k$ run over $\theta,\varphi$), one can easily get that $G^\mu_{t~|\mu}=G^\mu_{\theta~|\mu}=G^\mu_{\varphi~|\mu}=0$. The proof of $G^\mu_{r~|\mu}=0$ is somewhat subtle. By making use of the equations (\ref{geodesic spray t},\ref{geodesic spray theta},\ref{geodesic spray phi}), we find from (\ref{chern connection}) and $A^\rho_{\mu\nu}=A^i_{jk}$ that
\begin{equation}\label{chern connection1}
\Gamma^t_{rt}=\frac{B'}{2B},~~~\Gamma^\theta_{r\theta}=\Gamma^\varphi_{r\varphi}=\frac{1}{r}.
\end{equation}
Then, after a tedious calculation, one can check that the equation $G^\mu_{r~|\mu}=0$ indeed satisfy.
Following the sprite of general relativity, we propose that the gravitational field equation in the given Finsler spacetime (\ref{Schwarz like}) should be of the form
\begin{equation}\label{field equation}
G^\mu_\nu=8\pi_F G T^\mu_\nu,
\end{equation}
where $T^\mu_\nu$ is the energy-momentum tensor.
The volume of Finsler space \cite{Book Shen} is generally different with the one of Riemann geometry. We have used $4\pi_F$ to denote the volume of $\bar{F}$ in field equation (\ref{field equation}). Boundary condition gives a specific form of subspace $\bar{F}$. The volume of surface of subspace $\bar{F}$ can identify the value of $\pi_{F}$. For example, if we make $\bar{F}$ to be the form $F_{\rm FS}$ (\ref{Randers 1}), then $\pi_{F}=\pi$ according to discussion of the subsection B in Sec. II. The proposed form of field equation (\ref{field equation}) given here is not inconsistent with properties of the ansatz (\ref{Schwarz like}). The gravitational field equation (\ref{field equation}) valid for specific Finsler spacetime (\ref{Schwarz like}). A general field equation valid for arbitrary Finsler spacetime has been proposed by Pfeifer and Wohlfarth \cite{Pfeifer} or Vacaru \cite{Vacaru}. However, these field equations do not satisfy requirement of covariant conservation or make the energy-momentum tensor constructed on tangent bundle.

For simplicity, we set the energy-momentum tensor to be of the form
\begin{equation}
T^\mu_\nu={\rm diag}(\rho(r),-p(r),-p(r),-p(r)),
\end{equation}
where $\rho(r)$  and $p(r)$ are the energy density and pressure of the gravitational source, respectively. Then, by making use of equations (\ref{G t},\ref{G r},\ref{G theta}), we reduce the gravitational field equation to three independent equations
\begin{eqnarray}\label{conserve eq}
\frac{2p'}{\rho+p}&=&-\frac{B'}{B},\\
\label{field eq t}
\frac{A'}{rA^2}-\frac{1}{r^2A}+\frac{\lambda}{r^2}&=&8\pi_F G\rho,\\
\label{field eq r}
\frac{B'}{rAB}+\frac{1}{r^2A}-\frac{\lambda}{r^2}&=&8\pi_F G p.
\end{eqnarray}
The solution of equation (\ref{field eq t}) is given as
\begin{equation}\label{interior A}
A^{-1}=\lambda-\frac{2Gm(r)}{r},
\end{equation}
where $m(r)\equiv\int^r_0 4\pi_Fx^2\rho(x)dx$.
By making use of equation (\ref{interior A}), and plugging the equation (\ref{field eq r}) into (\ref{conserve eq}), we obtain that
\begin{equation}\label{OV eq}
-r^2p'=(\rho+p)(4\pi_F Gp r^3+Gm)\left(\lambda-\frac{2Gm}{r}\right)^{-1}.
\end{equation}
The equation (\ref{OV eq}) reduces to the famous Oppenheimer-Volkoff equation if Finsler spacetime $\bar{F}$ reduces to two dimensional Riemann sphere. Combining the modified Oppenheimer-Volkoff equation (\ref{OV eq}) with the equation of state, one can obtain the interior structure of gravitational source.

The interior solution (\ref{interior A}) should consistent with the exterior solution (\ref{solution A1}) at the boundary of the gravitational source. Therefore, we get
\begin{equation}
\label{boundary cond}
a\lambda=1.
\end{equation}
At last, combining the boundary condition (\ref{boundary cond}) with the requirement of Newtonian limit (\ref{Newtonian limit}), we get the exterior solution $B(r)$ and $A(r)$ as
\begin{eqnarray}\label{result B}
B(r)&=&1-\frac{2GM}{\lambda r},\\
\label{result A}
A(r)&=&\left(\lambda-\frac{2GM}{r}\right)^{-1}.
\end{eqnarray}

\section{Experimental constraint on Finslerian gravity}
\subsection{The motion of particles}
Plugging the equations of geodesic spray coefficients (\ref{geodesic spray t},\ref{geodesic spray r},\ref{geodesic spray theta},\ref{geodesic spray phi}) into the formula of geodesic equation (\ref{geodesic}), we obtain the geodesic equation of Finsler spacetime (\ref{Schwarz like})
\begin{eqnarray}\label{geodesic eq t}
0&=&\frac{d^2t}{d\tau^2}+\frac{B'}{B}\frac{dr}{d\tau}\frac{dt}{d\tau},\\
\label{geodesic eq r}
0&=&\frac{d^2r}{d\tau^2}+\frac{B'}{2A}\left(\frac{dt}{d\tau}\right)^2+\frac{A'}{2A}\left(\frac{dr}{d\tau}\right)^2-\frac{r}{A}\bar{F}^2\left(\frac{d\theta}{d\tau},\frac{d\varphi}{d\tau}\right),\\
\label{geodesic eq theta}
0&=&\frac{d^2\theta}{d\tau^2}+\frac{2}{r}\frac{dr}{d\tau}\frac{d\theta}{d\tau}+2\bar{G}^\theta,\\
\label{geodesic eq phi}
0&=&\frac{d^2\varphi}{d\tau^2}+\frac{2}{r}\frac{dr}{d\tau}\frac{d\varphi}{d\tau}+2\bar{G}^\varphi.
\end{eqnarray}
The solution of equation (\ref{geodesic eq t}) is
\begin{equation}\label{first eq motion}
B\frac{dt}{d\tau}=1,
\end{equation}
where we have set the integral constant to be $1$ by the normalization of $\tau$. Noticing that $y^\mu$ equals to $\frac{dx^\mu}{d\tau}$ along the geodesic, and by making use of the equations (\ref{geodesic eq theta},\ref{geodesic eq phi}), we find that
\begin{eqnarray}\label{derive sec eq motion}
\frac{d\bar{F}}{d\tau}=\frac{\partial \bar{F}}{\partial x^i}\frac{dx^i}{d\tau}+\frac{\partial \bar{F}}{\partial y^i}\frac{dy^i}{d\tau}=y^i\left(\frac{\partial \bar{F}}{\partial x^i}-\frac{2\bar{G}_i}{\bar{F}}\right)-2y^r\bar{F}=-\bar{F}\frac{d\ln r^2}{d\tau},
\end{eqnarray}
where $\bar{G}_i=\bar{g}_{ij}\bar{G}^j$ ($i,j$ run over $\theta,\varphi$), and we have used the fact that $\bar{F}$ is a homogenous function of $y$ of degree $1$ to derive the third equation of (\ref{derive sec eq motion}). The solution of equations (\ref{derive sec eq motion}) is given as
\begin{equation}\label{second eq motion}
r^2\bar{F}=J,
\end{equation}
where $J$ is an integral constant. By making use of the equations (\ref{first eq motion},\ref{second eq motion}), we find from geodesic equation (\ref{geodesic eq r}) that
\begin{equation}\label{derive thr eq motion}
\frac{d}{d\tau}\left(A\left(\frac{dr}{d\tau}\right)^2+\frac{J^2}{r^2}-\frac{1}{B}\right)=0.
\end{equation}
The solution of (\ref{derive thr eq motion}) is given as
\begin{equation}\label{third eq motion}
A\left(\frac{dr}{d\tau}\right)^2+\frac{J^2}{r^2}-\frac{1}{B}=A\left(\frac{dr}{d\tau}\right)^2+r^2\bar{F}^2-B\left(\frac{dt}{d\tau}\right)^2=-F^2,
\end{equation}
where we have used the equations (\ref{first eq motion},\ref{second eq motion}) to derive the second equation of (\ref{third eq motion}). The equation (\ref{third eq motion}) means that $F$ is constant along the geodesic.

Now, we have three solutions (\ref{first eq motion},\ref{second eq motion},\ref{third eq motion}) of the geodesic equations, the fourth one depends on the explicit form of two dimension Finsler space $\bar{F}$. However, we can still find some information of particle motion from the obtained solutions. Consider a particle move along radial direction, combining the equation (\ref{first eq motion}) with (\ref{second eq motion}), and by making use of the exterior solutions (\ref{result B},\ref{result A}) of $B(r)$ and $A(r)$, we obtain that
\begin{equation}\label{radial motion}
\frac{dr}{dt}=\sqrt{\lambda^{-1}-F^2\left(1-\frac{2GM}{\lambda r}\right)}\left(1-\frac{2GM}{\lambda r}\right).
\end{equation}
It is obvious from (\ref{radial motion}) that $\frac{dr}{dt}\rightarrow0$ while $r\rightarrow2GM/\lambda$. The modified Schwarzschild radius in Finsler spacetime (\ref{Schwarz like}) is
\begin{equation}
r_s=\frac{2GM}{\lambda}.
\end{equation}

\subsection{Classical tests}
The predictions of general relativity have been proved by four classical tests \cite{Weinberg}. If our spacetime is Finslerian, then it is necessary to test the validity of Finslerian gravity. In this subsection, our discussion are based on the following Finsler structure
\begin{equation}\label{Schwarz like 1}
F^2=\left(1-\frac{2GM}{r}\right)y^ty^t-\left(1-\frac{2GM}{r}\right)^{-1}y^r y^r-r^2F^2_{\rm FS},
\end{equation}
where the ``Finslerian sphere" $F_{\rm FS}$ is of the form (\ref{Randers 1}). The difference between Finslerian solution (\ref{Schwarz like 1}) and Schwarzschild solution lies on the ``Finslerian sphere" $F_{\rm FS}$. Two of four classical tests, namely, Radar echo delay and gravitational redshift relate to the radial motion of particle. One can also find from equation (\ref{radial motion}) that $\frac{dr}{dt}$ is the same with one in Schwarzschild spacetime while $\lambda=1$.

It is expected that the motion of particle in a bounded and unbounded orbit in Finsler spacetime is different from one in Schwarzschild spacetime. Since the ``Finslerian sphere" is non-reversible for $\varphi\rightarrow-\varphi$, it implies that particle moves along a given direction ($\varphi$) is different from its counter ($-\varphi$) in orbital motion. It is convenient to consider the orbit of particle confined to the equatorial plane $\theta=\pi/2$. Then, by making use of the metric of ``Finslerian sphere" (\ref{Randers 1}), the equation (\ref{second eq motion}) simplifies as
\begin{equation}\label{second eq motion1}
r^2\frac{d\varphi}{d\tau}=J_{\pm},
\end{equation}
where $J_{\pm}\equiv(1\pm\epsilon)J$, $J_{+}$ corresponds to a given direction and $J_{-}$ corresponds to its counter. Plugging the equation (\ref{second eq motion1}) into the solution of geodesic equation (\ref{third eq motion}), we get the equation for orbital motion
\begin{equation}\label{orbital eq}
(1\pm\epsilon)^2\left(\frac{dr}{d\varphi}\right)^2=\frac{r^4}{J^2}-\left(1-\frac{2GM}{r}\right)\left(\frac{F^2r^4}{J^2}+r^2\right).
\end{equation}
One should notice that $F$ is constant.  By reparameterizing curve parameter $\tau$, one can set $F$ equals 1 and 0 for massive and massless particles, respectively. The solution of the orbital equation (\ref{orbital eq}) gives deflection angle and precession of the orbit per revolution for unbounded and bounded orbit, respectively. Noticing that the orbital equation is the same with one in Schwarzschild spacetime if we transform $(1\pm\epsilon)\varphi$ into $\varphi$. Thus, recalling the results in Schwarzschild spacetime, in Finsler spacetime (\ref{Schwarz like 1}), we obtain the deflection angle for gravitational deflection of light,
\begin{equation}
\delta\alpha=(1\pm\epsilon)\frac{4GM}{\xi},
\end{equation}
where $\xi$ is the distance of closest approach. And the precession of the orbit per revolution is given as
\begin{equation}
\delta\varphi=(1\pm\epsilon)6\pi\frac{GM}{L},
\end{equation}
where $L$ is semi-latus rectum of the orbit.

The observations of very-long-baseline radio interferometry \cite{Stairs,Will} give constraint on the gravitational deflection of light in solar system. Its results yield a constraint on Finslerian parameter $\epsilon$. It is given as $\epsilon\sim2\times10^{-4}$. The observations of perihelion shift of Mercury \cite{Will} also give constraint on Finslerian parameter $\epsilon$. Its results yield $\epsilon<3\times10^{-3}$. The recent Michelson-Morley experiment carried through by M\"{u}ller et al. \cite{Muller} gives a precise limit on Lorentz invariance violation. Their experiment shows that the change of resonance frequencies of the optical resonators is of this magnitude $|\frac{\delta\omega}{\omega}|\sim10^{-16}$. This implies the Finslerian parameter $\epsilon$ should less than $10^{-16}$.

\section{Counterpart of Birkhoff's theorem}
In general relativity, the Birkhoff's theorem guarantees that the solution of vacuum field equation with spherical symmetry must be static. It means that its exterior solution must be Schwarzschild metric regardless of the evolution of gravitational source. It is necessary to investigate such issue in Finsler spacetime. In Riemann geometry, the spherical symmetry can be represented by a Riemannian sphere. Its metric is of the form (\ref{Riemann sphere}). The counterpart of Riemannian sphere in Finsler geometry is ``Finslerian sphere" (\ref{Randers 1}). In the above discussion, we gave a static solution of Finslerian vacuum field equation with the symmetry of ``Finslerian sphere". Now, we turn to investigate time-dependent Finsler spacetime with the symmetry of ``Finslerian sphere". Its Finsler structure is given as
\begin{equation}\label{Schwarz like t-depend}
F^2=B(r,t)y^ty^t-A(r,t)y^r y^r-r^2F^2_{FS}.
\end{equation}
Plugging the Finsler structure (\ref{Schwarz like t-depend}) into the formula (\ref{geodesic spray}), we obtain that
\begin{eqnarray}\label{geodesic spray t1}
G^t&=&\frac{B'}{2B}y^t y^r+\frac{\dot{B}}{4B}y^t y^t+\frac{\dot{A}}{4B}y^r y^r,\\
\label{geodesic spray r1}
G^r&=&\frac{A'}{4A}y^r y^r+\frac{B'}{4A}y^t y^t-\frac{r}{2A}F_{FS}^2+\frac{\dot{A}}{2A}y^t y^r,\\
\label{geodesic spray theta1}
G^\theta&=&\frac{1}{r}y^\theta y^r+G_{FS}^\theta,\\
\label{geodesic spray phi1}
G^\varphi&=&\frac{1}{r}y^\varphi y^r+G_{FS}^\varphi,
\end{eqnarray}
where the dot denotes the derivative with respect to $t$, and the $\bar{G}$ is the geodesic spray coefficients derived by $F_{FS}$. Plugging the geodesic coefficients (\ref{geodesic spray t1},\ref{geodesic spray r1},\ref{geodesic spray theta1},\ref{geodesic spray phi1}) into the formula of Ricci scalar (\ref{Ricci scalar}), we obtain
\begin{eqnarray}
F^2Ric=&&\left[\frac{B''}{2A}-\frac{B'}{4A}\left(\frac{A'}{A}+\frac{B'}{B}\right)+\frac{B'}{r A}-\frac{\ddot{A}}{2A}+\frac{\dot{A}^2}{4A^2}+\frac{\dot{A}\dot{B}}{4AB}\right]y^ty^t\nonumber\\
&+&\left[-\frac{B''}{2B}+\frac{B'}{4B}\left(\frac{A'}{A}+\frac{B'}{B}\right)+\frac{A'}{r A}+\frac{\ddot{A}}{2B}-\frac{\dot{A}\dot{B}}{4B^2}-\frac{\dot{A}^2}{4AB}\right]y^r y^r\nonumber\\
\label{Ricci scalar time}
&+&\frac{2\dot{A}}{rA}y^t y^r+\left[1-\frac{1}{A}+\frac{r}{2A}\left(\frac{A'}{A}-\frac{B'}{B}\right)\right]F_{FS}^2~.
\end{eqnarray}
The vacuum field equation $Ric=0$ means that $\frac{2\dot{A}}{rA}=0$. It tells us that $A$ is time-independent. This fact shows that all time derivatives drop out of the equation (\ref{Ricci scalar time}), and it becomes identical with the one in static case. Following the discussion for static case, we obtain that
\begin{eqnarray}
B&=&f(t)\left(1-\frac{2GM}{r}\right),\\
A&=&\left(1-\frac{2GM}{r}\right)^{-1}.
\end{eqnarray}
The function $f(t)$ can be made to equal 1 by defining a new time coordinate
\begin{equation}
t'=\int^t \sqrt{f(t)}dt.
\end{equation}
Now, the Finsler structure (\ref{Schwarz like t-depend}) is entirely time-independent and identical with the static solution (\ref{Schwarz like 1}). Thus, the counterpart of the Birkhoff's theorem exists in Finslerian gravity. Unlike the requirement of spherical symmetry in general relativity, it shows that the Finslerian gravitational field with the symmetry of ``Finslerian sphere" in vacuum must be static, and its metric is of the form (\ref{Schwarz like 1}).

\section{Conclusions and Remarks}
In view of geodesic deviation equation, the vacuum field equation $Ric=0$ in Finsler spacetime implies that the geodesic rays are parallel to each other. The geometric invariant property of Ricci scalar implies that the vacuum field equation is insensitive to the connection, which is an essential physical requirement. Starting from the ansatz (\ref{Schwarz like}), we have found an exact solution of vacuum field equation (\ref{solution B1},\ref{solution A1}).

A general gravitational field equation in Finsler spacetime is still to be completed. However, we have found that the proposed form of field equation (\ref{field equation}) given here is not inconsistent with properties of the ansatz (\ref{Schwarz like}). And, we have proved that the Finslerian covariant derivative of the geometrical part of the gravitational field equation is conserved. It is obvious that the gravitational field equation (\ref{field equation}) returns to the vacuum field equation while the energy-momentum tensor vanishes. We have found an interior solution of the gravitational field equation (\ref{field equation}). The interior solution (\ref{interior A}) consistent with the exterior solution (\ref{solution A1}) at the boundary of the gravitational source. And we required that the exterior solution should return to Newton's gravity. The two boundary conditions constrain that the exterior solution should be the form as (\ref{result B},\ref{result A}). One should notice that Schwarzschild metric is also a solution of $Ric=0$. There is a boundary condition which distinguish Finslerian solution from Schwarzschild solution. It is violation of Lorentz symmetry while the Finsler spacetime (\ref{Schwarz like}) has no gravitational source $M=0$. For example, we make subspace $\bar{F}$ to be a ``Finslerian sphere" $F_{FS}$. Then, the exterior metric of vacuum field solution was given as
\begin{eqnarray}\label{final metric}
F^2=&&\left(1-\frac{2GM}{r}\right)y^ty^t-\left(1-\frac{2GM}{r}\right)^{-1}y^r y^r\nonumber\\
&-&r^2\left(\frac{\sqrt{(1-\epsilon^2\sin^2\theta) y^\theta y^\theta+\sin^2\theta y^\varphi y^\varphi}-\epsilon\sin^2\theta y^\varphi}{1-\epsilon^2\sin^2\theta}\right)^2.
\end{eqnarray}
The metric (\ref{final metric}) is no other than the Schwarzschild metric except for the change from Riemannian sphere to ``Finslerian sphere" (\ref{Randers 1}). We have presented three solutions (\ref{first eq motion},\ref{second eq motion},\ref{third eq motion}) of geodesic equations of the metric (\ref{final metric}). The fourth one depends on the geodesic equation of the ``Finslerian sphere" (\ref{Randers 1}). The geometrical properties of `Finslerian sphere" (\ref{Randers 1}) are as follows. It is non-reversible for $\varphi\rightarrow-\varphi$, it has two close geodesics locate at $\theta=\frac{\pi}{2}$ with length $L_{\pm}=2\pi(1\pm\epsilon)^{-1}$, its volume that is the surface volume of unit ``Finslerian sphere" equals to $4\pi$, it only have one independent Killing vector $V^\varphi=\rm constant$.

We have investigated the motion of particles in Finsler spacetime (\ref{Schwarz like}). The solution of geodesic equations are shown. Taking Finsler spacetime (\ref{Schwarz like 1}) as a example, we have shown the experimental constraint on Finslerian parameter $\epsilon$. In solar system, the celestial observations require $\epsilon<10^{-4}$. And, the recent Michelson-Morley experiment requires $\epsilon<10^{-16}$. It is expected that Finslerian spacetime may have unignorable effect in cosmological scale. In general relativity, as a special case of gravitational lensing, Einstein ring has a symmetric structure. However, if the spacetime is Finslerian, then one may observed Einstein ring with asymmetric structure.

The counterpart of Birkhoff's theorem exist in Finslerian vacuum. It shows that the Finslerian gravitational field with the symmetry of ``Finslerian sphere" in vacuum must be static, and its metric is of the form (\ref{Schwarz like 1}).

The Schwarzschild spacetime will return to Minkowski spacetime if there is no gravitational source. As for the Finslerian vacuum spacetime (\ref{final metric}), if there is no gravitational source, namely, $M=0$, the metric reduces to
\begin{eqnarray}\label{M=0}
F^2=&&y^ty^t-y^r y^r-r^2\left(\frac{\sqrt{(1-\epsilon^2\sin^2\theta) y^\theta y^\theta+\sin^2\theta y^\varphi y^\varphi}-\epsilon\sin^2\theta y^\varphi}{1-\epsilon^2\sin^2\theta}\right)^2.
\end{eqnarray}
According to the formula (\ref{Ricci scalar1}), the Ricci scalar or Ricci tensor of the metric (\ref{M=0}) equal to $0$. This fact holds even for three dimensional subspace of the metric (\ref{M=0}). However, the metric (\ref{M=0}) or its spatial part is not a flat Finsler spacetime. The Finsler spacetime is a flat one \cite{Book by Bao} if and only if $Ric=0$ and the geodesic spray coefficients $G^\mu$ is quadratic in $y$. This fact is quite different with Riemann geometry. It is well known that three dimensional Riemann space is flat while its Ricci tensor equal to $0$. Nevertheless, even the space part of the metric (\ref{M=0}) is not a flat Finsler space.

In Riemann spacetime without torsion, at any fixed point, one can erect a local coordinate system such that the metric is Minkowskian. One necessary condition of this statement is the Riemann metric is quadric. However, this necessary condition does not hold in a general Finsler metric. Therefore, Finsler spacetime is not locally isometric to a Minkowski spacetime. One deduction of it is the speed of light is not locally isotropic. The propagation of light obeys $F=0$. One can find from the local metric that the radial speed of light equals to $1$. And the non-radial speed of light satisfy
\begin{equation}
c_\theta^2+(c_\varphi-\epsilon \sin\theta)^2=1,
\end{equation}
where $c_\theta\equiv\frac{d\theta}{dt}$ and $c_\varphi\equiv\frac{d\varphi}{dt}\sin\theta$.

The Schwarzschild radius form an event horizon in Schwarzschild spacetime. The Schwarzschild solution can be maximally extended by Kruskal extension. The coordinate transformation between Schwarzschild metric and Kruskal metric is only related to $r$ and $t$. Therefore, one can also get a maximally extended Finslerian vacuum solution (\ref{final metric}) by Kruskal's method.

\vspace{1cm}
\begin{acknowledgments}
We would like to thank Prof. C.-G. Huang for useful discussions. Project 11375203 and 11305181 supported by NSFC.
\end{acknowledgments}

\end{document}